\documentclass[prl,twocolumn,showpacs,superscriptaddress,floatfix]{revtex4}

\usepackage{graphicx}  
\usepackage{makeidx}
\usepackage{colordvi}
\usepackage{dcolumn}
\usepackage{amsmath}

\RequirePackage{xspace}

\usepackage{relsize}
\def\babar{\mbox{\slshape B\kern-0.1em{\smaller A}\kern-0.1em
    B\kern-0.1em{\smaller A\kern-0.2em R}}}
\def\epem       {\ensuremath{e^+e^-}\xspace}
\def\qqbar {\ensuremath{q\overline q}\xspace}
\def\ccbar {\ensuremath{c\overline c}\xspace}
\def\piz   {\ensuremath{\pi^0}\xspace}
\def\pip   {\ensuremath{\pi^+}\xspace}
\def\pim   {\ensuremath{\pi^-}\xspace}

\def\Kbar  {\kern 0.2em\overline{\kern -0.2em K}{}\xspace}

\def\Km    {\ensuremath{K^-}\xspace}
\def\KS    {\ensuremath{K^0_{\scriptscriptstyle S}}\xspace}
\def\Dbar    {\kern 0.2em\overline{\kern -0.2em D}{}\xspace}

\def\Dz      {\ensuremath{D^0}\xspace}
\def\Dzb     {\ensuremath{\Dbar^0}\xspace}
\def\Dp      {\ensuremath{D^+}\xspace}
\def\Dm      {\ensuremath{D^-}\xspace}

\def\DpDm    {\ensuremath{\Dp {\kern -0.16em \Dm}}\xspace}
\def\Dstar   {\ensuremath{D^*}\xspace}

\def\Dstarzb {\ensuremath{\Dbar^{*0}}\xspace}
\def\Dstarp  {\ensuremath{D^{*+}}\xspace}
\def\Dstarm  {\ensuremath{D^{*-}}\xspace}

\def\Dss     {\ensuremath{D^{*+}_s}\xspace}

\def\Bbar    {\kern 0.18em\overline{\kern -0.18em B}{}\xspace}

\def\BB      {\ensuremath{B\Bbar}\xspace}
\def\Bz      {\ensuremath{B^0}\xspace}
\def\Bzb     {\ensuremath{\Bbar^0}\xspace}
\def\BzBzb   {\ensuremath{\Bz {\kern -0.16em \Bzb}}\xspace}
\def\Bu      {\ensuremath{B^+}\xspace}

\def\Bp      {\ensuremath{\Bu}\xspace}

\def\Y#1S{\ensuremath{\Upsilon{(#1S)}}\xspace}% no space before {...}!
\def\FourS {\Y4S}

\def\Bztodstdst {\ensuremath{\Bz \to \Dstarp \Dstarm}\xspace}

\def\upsbb   {\ensuremath{\FourS \to \BB}\xspace}
\def\mes        {\mbox{$m_{\rm ES}$}\xspace}
\newcommand{\mev}{\ensuremath{\mathrm{\,Me\kern -0.1em V}}\xspace}
\newcommand{\gevc}{\ensuremath{{\mathrm{\,Ge\kern -0.1em V\!/}c}}\xspace}
\newcommand{\mevc}{\ensuremath{{\mathrm{\,Me\kern -0.1em V\!/}c}}\xspace}
\newcommand{\gevcc}{\ensuremath{{\mathrm{\,Ge\kern -0.1em V\!/}c^2}}\xspace}
\newcommand{\mevcc}{\ensuremath{{\mathrm{\,Me\kern -0.1em V\!/}c^2}}\xspace}
\def\to                 {\ensuremath{\rightarrow}\xspace}
\newcommand{\stat}{\ensuremath{\mathrm{(stat)}}\xspace}
\newcommand{\syst}{\ensuremath{\mathrm{(syst)}}\xspace}
\def\pep2{PEP-II}

\def\CP                {\ensuremath{C\!P}\xspace}
\def\stwob{\ensuremath{\sin\! 2 \beta   }\xspace}

\def\deltat{\ensuremath{{\rm \Delta}t}\xspace}
\def\deltamd{\ensuremath{{\rm \Delta}m_d}\xspace}
\def\ps         {\ensuremath{{\rm \,ps}}\xspace}  %% picosecond

\newcommand{\progtp}    [1]  {{Prog.\ Theor.\ Phys.\ {\bf #1}}}

\begin{document}  
\begin{flushleft}
\babar-PUB-07/048 \\
SLAC-PUB-12725\\
\end{flushleft}

\title{
{\large  \boldmath
Improved Measurement of Time-Dependent \CP Asymmetries and the \\
\CP-Odd Fraction in the Decay \Bztodstdst}
}
 
%% author list as of 05-Jul-2007 (565 authors)
%
\author{B.~Aubert}
\author{M.~Bona}
\author{D.~Boutigny}
\author{Y.~Karyotakis}
\author{J.~P.~Lees}
\author{V.~Poireau}
\author{X.~Prudent}
\author{V.~Tisserand}
\author{A.~Zghiche}
\affiliation{Laboratoire de Physique des Particules, IN2P3/CNRS et Universit\'e de Savoie, F-74941 Annecy-Le-Vieux, France }
\author{J.~Garra~Tico}
\author{E.~Grauges}
\affiliation{Universitat de Barcelona, Facultat de Fisica, Departament ECM, E-08028 Barcelona, Spain }
% ...
\collaboration{The \babar\ Collaboration}
\noaffiliation

\begin{abstract} 
We present an updated measurement of the \CP-odd fraction and the
time-dependent \CP asymmetries in the decay $B^0 \rightarrow
D^{*+}D^{*-}$ using $(383\pm 4) \times 10^{6} \BB$ pairs collected
with the \babar\ detector.  We determine the \CP-odd fraction to be
$0.143 \pm 0.034\stat \pm 0.008\syst$.  The time-dependent \CP
asymmetry parameters are determined to be $C_+ = -0.05\pm 0.14\stat
\pm 0.02\syst$ and $S_+ = -0.72 \pm 0.19\stat \pm 0.05\syst$.
The non-zero value of the measured $S_+$ indicates the evidence of 
CP violation at the $3.7\,\sigma$ confidence level.
\end{abstract}
 
% PACS, the Physics and Astronomy Classification Scheme.
  
\maketitle

In the Standard Model (SM), \CP violation is described by a single
complex phase in the Cabibbo-Kobayashi-Maskawa (CKM) quark-mixing
matrix, $V$~\cite{CKM}. Measurements of \CP asymmetries by the
\babar~\cite{Aubert:2002ic} and Belle~\cite{Abe:2002px} collaborations
have firmly established this effect in the $b \to (\ccbar) s$
charmonium decays~\cite{conjugate} and precisely determined the
parameter \stwob, where $\beta$ is ${\rm arg}[-V_{\rm cd}V^*_{\rm
cb}/V_{\rm td}V^*_{\rm tb}]$.  The amplitude of the decay \Bztodstdst
is dominated by a tree-level, color-allowed $b\to \ccbar d$
transition.  Within the framework of the SM, the \CP asymmetry of
\Bztodstdst is equal to \stwob when the correction due to penguin
diagram contributions is neglected. The penguin-induced correction to
the \CP asymmetry, estimated in models based on the factorization
approximation and heavy quark symmetry, is predicted to be about
2\%~\cite{Pham:1999fy}, while contributions from non-SM processes may lead
to a large correction~\cite{Grossman:1996ke}.  Such a deviation in
the \stwob measurement from that of the $\Bz\to(\ccbar)K^{(*)0}$
decays would be evidence of physics beyond the SM.

Studies of \CP violation in $\Bz\to D^{(*)\pm}D^{(*)\mp}$ transitions
have been carried out by both the \babar\ and Belle
collaborations. Most recently, the Belle collaboration reported
evidence of large direct \CP violation in $\Bz\to\Dp\Dm$ where
$C_{\Dp\Dm} = -0.91\pm 0.23\pm 0.06$~\cite{Fratina:2007zk}, in
contradiction to the SM expectation. However, a large direct \CP
violation has not been observed in this channel by
\babar~\cite{Aubert:2007pa}, nor in previous measurements with
\Bztodstdst decays that involve the same quark-level weak
decay~\cite{Aubert:2005rn,Miyake:2005qb}.

The \Bztodstdst decay proceeds through the \CP-even $S$ and $D$ waves
and through the \CP-odd $P$ wave.  In this Letter, we present an
improved measurement of the \CP-odd fraction $R_\perp$ 
based on a time-integrated one-dimensional angular
analysis. We also present an improved measurement of the
time-dependent \CP asymmetry, obtained from a
combined analysis of time-dependent flavor-tagged decays and the
one-dimensional angular distribution of the decay products.

The data used in this analysis comprise $(383\pm 4)\times 10^{6}$
\upsbb decays collected with the \babar\ detector~\cite{Aubert:2001tu}
at the PEP-II asymmetric-energy \epem storage rings.  We use a Monte
Carlo (MC) simulation based on GEANT4~\cite{Agostinelli:2002hh} to
validate the analysis procedure and to study the relevant backgrounds.

We select \Bztodstdst candidates from oppositely charged pairs of
\Dstar mesons.  The \Dstarp is reconstructed in its decays to
$\Dz\pip$ and $\Dp\piz$.  We reconstruct candidates for \Dz and \Dp
mesons in the modes $\Dz\to\Km\pip$, $\Km\pip\piz$, $\Km\pip\pip\pim$,
$\KS\pip\pim$ and $\Dp\to\Km\pip\pip$.  We reject the \Bz candidates
for which both \Dstar mesons decay to $D\piz$ because of the smaller
branching fraction and larger backgrounds.  To suppress the
$\epem\to\qqbar \;(q=u,d,s,\,{\rm{and}}\; c)$ continuum background, we
require the ratio of the second and zeroth order Fox-Wolfram
moments~\cite{Fox:1978vu} to be less than 0.6.

For each \Bztodstdst candidate, we construct a likelihood function
$\mathcal{L}_{\rm{mass}}$ from the masses and mass uncertainties of
the $D$ and \Dstar candidates~\cite{Aubert:2006ia}.  In this
likelihood, the $D$ mass resolution is modeled by a Gaussian function whose
variance is determined candidate-by-candidate from the mass
uncertainty resulting from a vertex fit of the $D$ meson decay products.
The $\Dstar-D$ mass
difference resolution is modeled by the sum of two Gaussian
distributions whose parameters are determined from simulated events.
The maximum allowed values of $-\ln\mathcal{L}_{\rm{mass}}$ and $|\Delta E|\equiv
|E_B^*-E_{\rm{beam}}|$, the difference between the \Bz candidate
energy $E_B^*$ and the beam energy $E_{\rm{beam}}$ in the \FourS
frame, are optimized separately for each final state using simulated events to
obtain the highest expected signal significance.

We include candidates with an energy-substituted mass, $m_{\rm
ES}\equiv \sqrt{E^2_{\rm beam}-p^{*2}_B}$, greater than $5.23\,\gevcc$,
where $p^*_B$ is the $B^0$ candidate momentum in the \FourS frame.
On average, we have 1.8 \Bz candidates per event in data after all the
selection requirements.  In cases where more than one candidate is
reconstructed in an event, the candidate with the smallest value of
$-\ln\mathcal{L}_{\rm{mass}}$ is chosen.  Studies using MC samples
show that this procedure results in the selection of the correct \Bz
candidate more than 95\,\% of the time.

The total probability density function (PDF) of the $m_{\rm{ES}}$
distribution is the sum of the signal and background components.  The
signal PDF is modeled by a Gaussian function and the combinatorial
background is described by a threshold
function~\cite{Albrecht:1990cs}. Studies based on MC simulation
show that there is a small
peaking background from $\Bp\to\Dstarzb\Dstarp$ in which a $\Dzb$
originating from a \Dstarzb\ decay is combined with a random soft
\pim\ to form a \Dstarm\ candidate. This background is
described by the same PDF as the signal, and its fraction with respect to
the signal yield is fixed to $(1.8\pm1.8)\%$, as determined in MC
simulation.  
An unbinned maximum likelihood (ML) fit to the $m_{\rm{ES}}$ distribution
yields $617\pm 33\stat$ signal events, where the mean and width of the 
signal Gaussian function and the threshold function shape parameters 
are allowed to vary in the fit.  
The signal purity in the region of $m_{\rm{ES}}\ge 5.27\,\gevcc$
is approximately 65\,\%.

\begin{figure}[hbt]
\includegraphics[clip=true,width=0.9\columnwidth]{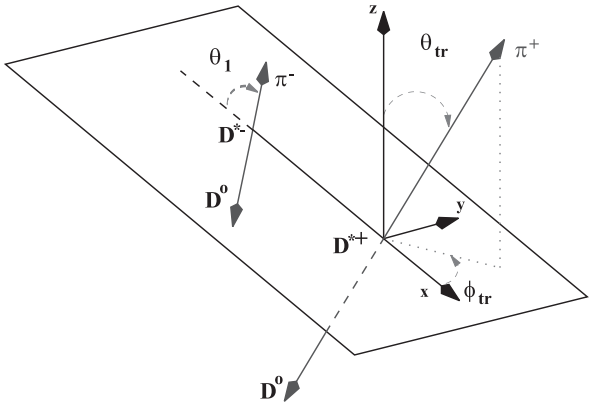}
\caption{\label{fig:trans}Depiction of the $B^0 \to \Dstarp\Dstarm$
decay in the transversity basis.  The three transversity angles are
defined in the text.}
\end{figure}

Following \cite{Dunietz:1990cj}, we define three
angles depicted in Figure~\ref{fig:trans} within the transversity
framework: the angle $\theta_1$ between the momentum of the slow pion
from the \Dstarm and the direction opposite the \Dstarp flight in the
\Dstarm rest frame; the polar angle $\theta_{\rm tr}$ and azimuthal
angle $\phi_{\rm tr}$ of the slow pion from the \Dstarp evaluated in
the \Dstarp rest frame, where the coordinate system is defined with the
$z$ axis normal to the $D^{*-}$ decay plane and the
$x$ axis opposite to the $D^{*-}$ momentum.

The time-dependent angular distribution of the decay products is given
in Ref.~\cite{penguin}.  Taking into account the detector efficiency
as a function of the transversity angles and integrating over the
decay time and the angles $\theta_1$ and $\phi_{\rm tr}$, we obtain a
one-dimensional differential decay rate:
\begin{eqnarray}
\frac{1}{\Gamma}  \frac{d \Gamma}{d \cos \theta_{\rm tr}} & = &
  \frac{9}{32\pi} \left\{ (1-R_\perp) \sin^2 \theta_{\rm tr}  \right. \nonumber \\
& \times &  \left[ \frac{1+\alpha}{2} I_{0}(\cos\theta_{\rm tr}) +
\frac{1-\alpha}{2} I_{\parallel}(\cos\theta_{\rm tr})
\right] \nonumber \\
& + &  \left. 2 R_\perp \cos^2 \theta_{\rm tr}
\times I_{\perp}(\cos\theta_{\rm tr}) \right\},
\label{AngDisArt}
 \end{eqnarray}
where
$R_\perp = |A_\perp|^2 / (|A_0|^2 + |A_\parallel|^2 + |A_\perp|^2)$,
$\alpha = (|A_{0}|^2  - |A_{\parallel}|^2) / (|A_0|^2 + |A_{\parallel}|^2)$,
$A_0$ is the amplitude for longitudinally polarized $D^*$ mesons,
$A_{\parallel}$ and $A_{\perp}$ are the amplitudes for parallel and
perpendicular transversely polarized $D^*$ mesons.  The three
efficiency moments $I_k(\cos\theta_{\rm tr})$, where $(k=0, \parallel, \perp)$, 
are defined as
\begin{eqnarray}
\displaystyle
I_{k}(\cos\theta_{\rm tr}) = \int \! d \! \cos \theta_1 \, d \phi_{\rm tr}
\; g_k(\theta_1, \phi_{\rm tr}) \,\varepsilon(\theta_1,\theta_{\rm tr},\phi_{\rm tr}),
\label{moments}
\end{eqnarray}
where $g_0 = 4\cos^2\theta_1 \cos^2\phi_{\rm tr}$, $g_{||} =
2\sin^2\theta_1 \sin^2\phi_{\rm tr}$, $g_{\perp} = \sin^2\theta_1 $,
and $\varepsilon$ is the overall detector efficiency.  The efficiency
moments are parameterized as second-order even polynomials of
$\cos\theta_{\rm tr}$ with parameter values determined from the MC
simulation. In fact, the three $I_{k}$ functions deviate only slightly from a
constant, making the decay distribution (Eq.~\ref{AngDisArt}) nearly
independent of the amplitude asymmetry $\alpha$.

\begin{figure}[bth]
\includegraphics[width=\columnwidth]{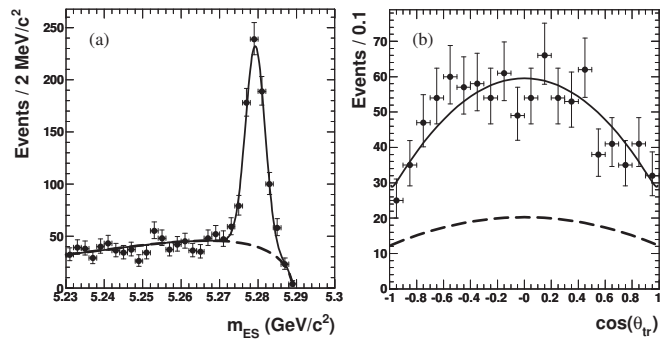}
\caption{ Measured distribution of $m_{\rm ES}$ (a) and of
$\cos\theta_{\rm tr}$ in the region $\mes > 5.27\,\gevcc$ (b).
The solid line is the projection of the fit result.  The dotted line
represents the background component.  }
\label{fig:datafit_rt}
\end{figure}

The \CP-odd fraction $R_\perp$ is measured in a simultaneous unbinned ML fit 
to the $\cos\theta_{\rm tr}$ and the $m_{\rm ES}$ distributions shown in
Figure~\ref{fig:datafit_rt}.  The background in the $\cos\theta_{\rm tr}$
distribution is modeled as an even, second-order polynomial, 
while the signal PDF is given by
Eq.~\ref{AngDisArt}. The finite detector resolution of the
$\theta_{\rm tr}$ measurement is modeled by the sum of three Gaussian
functions plus a small tail component that accounts for misreconstructed 
events, where all the parameters are
fixed to the values determined in the MC simulation.  The resolution
function is convolved with the signal PDF in the maximum likelihood
fit. We categorize events into three
types: $\Dstarp\Dstarm \to (\Dz\pip,\Dzb\pim)$, $(\Dz\pip,\Dm\piz)$,
and $(\Dp\piz,\Dzb\pim)$, each with different signal-fraction
parameters in the likelihood fit. Their efficiency moments and
$\cos\theta_{\rm tr}$ resolutions are separately determined from the
MC simulation. The other parameters, determined in the likelihood fit,
are the $\cos\theta_{\rm tr}$ background-shape parameter, three
$m_{\rm ES}$ parameters (width and mean of the signal Gaussian, and
the threshold function shape parameter), as well as $R_\perp$.

After fitting to data and taking into account possible systematic
uncertainties, we find 
\begin{equation}
\displaystyle
R_\perp = 0.143 \pm 0.034\stat \pm 0.008\syst\,.
\end{equation}
Figure~\ref{fig:datafit_rt} shows the
projections of the data and the fit result onto $m_{\rm ES}$ and
$\cos\theta_{\rm tr}$.

In the fit described above, the value of $\alpha$, the asymmetry
between the two \CP-even amplitudes in the transversity framework, is
fixed to zero.  We estimate the corresponding systematic uncertainty
by varying its value from $-1$ to $+1$ and find negligible change
(0.003) in the fitted value of $R_\perp$.  Other systematic
uncertainties arise from varying fixed parameters within their
errors: the parameterization of the angular resolution (0.006), the
determination of the efficiency moments (0.004), and the background
parameterization (0.004). The total systematic uncertainty on
$R_\perp$ is 0.008.

We perform a combined analysis of the $\cos\theta_{\rm
tr}$ distribution and its time dependence to extract the
time-dependent \CP asymmetry, using the event sample described
previously. We use information from the other $B$ meson ($B_{\rm
tag}$) in the event to tag the initial flavor of the fully
reconstructed \Bztodstdst candidate ($B_{\rm rec}$). The multivariate
flavor tagging algorithm is described in detail
elsewhere~\cite{Aubert:2004zt}.  We define six mutually exclusive
tagging categories in order of expected tag purity from lepton to
hadron, which includes kaon and pion tags. The total effective tagging
efficiency of this algorithm is $(30.5\pm 0.4)\,\%$.

The decay rate $f_+ (f_-)$ for a neutral $B$ meson accompanied by a
$B^{0} (\Bzb)$ tag is given by
\begin{eqnarray}
f_\pm(\deltat,\cos\theta_{\rm tr}) \propto
{\rm e}^{ - | \deltat |/\tau_{B^0} }
 \Bigl\{ G(1\mp\Delta\omega) \pm (1-2\omega)\nonumber \\
 \left[
 F\sin{ (\deltamd  \deltat) }
- H\cos{ (\deltamd  \deltat) }  \right]  \Bigr\},
 \label{eq:sincos}
\end{eqnarray}
where $\Delta t = t_{\rm rec} - t_{\rm tag}$ is the difference between
the proper decay time of the $B_{\rm rec}$ and $B_{\rm tag}$ mesons,
$\tau_{\Bz}=(1.530\pm 0.009)\,\ps$ is the \Bz lifetime, and 
$\deltamd=(0.507\pm 0.005)\, \ps^{-1}$ is the mass difference
between the \Bz-\Bzb mass eigenstates~\cite{Yao:2006px}.
The average mistag probability $\omega$ describes the effect of
incorrect tags, and $\Delta\omega$ is the difference between the
mistag rate for $\Bz$ and $\Bzb$. The $G$, $F$ and $H$ coefficients
are defined as:
\begin{eqnarray}
G &=&  (1-R_\perp) \sin^2\theta_{\rm tr}
   +    2  R_\perp \cos^2\theta_{\rm tr},\nonumber \\
F &=&  (1-R_\perp)S_+ \sin^2\theta_{\rm tr}-
2R_\perp S_\perp\cos^2\theta_{\rm tr},\\
H &=&  (1-R_\perp)C_+ \sin^2\theta_{\rm tr}+
2R_\perp C_\perp\cos^2\theta_{\rm tr},\nonumber 
\end{eqnarray}
where we allow the three transversity amplitudes to have different
$\lambda_k=(q/p)(\bar{A}_k/A_k)$ $(k=0, \parallel,
\perp)$~\cite{penguin} due to possibly different penguin-to-tree
amplitude ratios, and define the \CP asymmetry parameters
$C_k=(1-|\lambda_k|^2)/(1+|\lambda_k|^2)$,
$S_k=2\mathcal{I}m(\lambda_k)/(1+|\lambda_k|^2)$.  Here, we also define:
\begin{equation}
C_+=\frac{C_\parallel |A_\parallel|^2+C_0|A_0|^2}
{|A_\parallel|^2+|A_0|^2},
%\nonumber \\
S_+=\frac{S_\parallel |A_\parallel|^2+S_0|A_0|^2}
{|A_\parallel|^2+|A_0|^2}.
\end{equation}
In the absence of penguin contributions, we expect that
$C_0=C_\parallel=C_\perp=0$, and
$S_0=S_\parallel=S_\perp=-\stwob$~\cite{Pham:1999fy}.

In Eq.~\ref{eq:sincos}, the small detector efficiency effects are not
taken into account and instead are absorbed into the value of
$R_\perp$, which is allowed to vary in the final fit.  Any bias in
the resulting values of $C_+$, $C_\perp$, $S_+$, and $S_\perp$ is
below the sensitivity of our MC validation sample and is accounted for
in the MC statistics systematic.  Hence, a dedicated method to correct
for detector efficiency is not required. However, the ``effective''
value of $R_\perp$ obtained in this way is not identical to the value
measured from the time-integrated analysis that includes the
efficiency correction.  This approach incorporates the
uncertainty in $R_\perp$ directly into the determination of the 
\CP parameters in the ML fit.

The technique used to measure the \CP asymmetry is analogous to that
used in \babar\ measurements as described in
Ref.~\cite{Aubert:2004zt,Aubert:2002rg}.  We calculate the time
interval \deltat between the two $B$ decays from the measured
separation $\Delta z$ between the decay vertices of $B_{\rm rec}$ and
$B_{\rm tag}$ along the collision ($z$) axis~\cite{Aubert:2002rg}. The
$z$ position of the $B_{\rm rec}$ vertex is determined from the
charged daughter tracks. The $B_{\rm tag}$ decay vertex is determined
by fitting charged tracks not belonging to the $B_{\rm rec}$ candidate
to a common vertex, employing constraints from the beam spot location
and the $B_{\rm rec}$ momentum~\cite{Aubert:2002rg}.  Only events with
a $\Delta t$ uncertainty less than $2.5\,\mbox{ps}$ and a measured
$|\Delta t|$ less than $20\,\mbox{ps}$ are accepted.  We perform a
simultaneous unbinned ML fit to the $\cos\theta_{\rm tr}$, $\Delta t$,
and $m_{\rm ES}$ distributions to extract the \CP asymmetry. The
signal PDF in $\theta_{\rm tr}$ and $\Delta t$ is given by
Eq.~\ref{eq:sincos}.  The signal mistag probability and the difference 
between the mistag rate for $\Bz$ and $\Bzb$ are determined 
for each tagging category from
a large sample of neutral $B$ decays to flavor eigenstates, $B_{\rm flav}$.  
In the likelihood fit, the expression in Eq.~\ref{eq:sincos}
is convolved with an empirical $\Delta t$ resolution function
determined from the $B_{\rm flav}$ sample. The $\theta_{\rm tr}$
resolution is accounted for in the same way as described previously.

Our increased statistics allows for better treatment of the background
in this analysis.  The background $\Delta t$ distributions are
parameterized by an empirical description that includes components
that have zero lifetime, and that have an effective lifetime
similar to the signal.  The lifetime of the second component and its
relative fraction are allowed to vary in the likelihood fit.  We also
allow the lifetime component to have free effective \CP asymmetry parameters, 
$C_{\rm eff}$ and $S_{\rm eff}$, for each tagging
category to take into account a possible difference in mistag rates in
the background.  The background shape in $\theta_{\rm tr}$ is modeled
by an even, second-order polynomial in $\cos\theta_{\rm tr}$, as in
the time-integrated angular analysis.
 
\begin{figure}[htb]
\includegraphics[width=0.8\columnwidth]{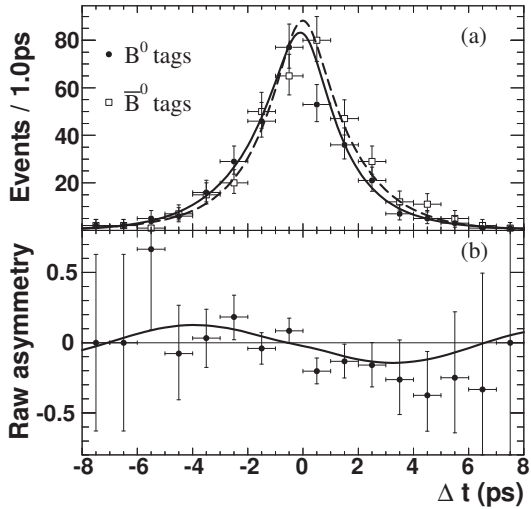}
\caption{The distribution in $\Delta t$ of the yield in the region
$\mes > 5.27\,\gevcc$ for \Bz (\Bzb) tagged candidates (a) and the raw
asymmetry $(N_{\Bz}-N_{\Bzb})/(N_{\Bz}+N_{\Bzb})$, as functions of
\deltat (b).  In (a), the solid (dashed) curves represent the fit to
the data for \Bz (\Bzb) tags.  }
\label{fig:datafit_cp}
\end{figure}

From our fit to data we determine
\begin{eqnarray}
C_+&=&-0.05\pm 0.14\stat\pm 0.02\syst,\nonumber\\
C_\perp&=&~~0.23\pm 0.67\stat\pm 0.10\syst,\nonumber\\
S_+&=&-0.72\pm 0.19\stat\pm 0.05\syst,\nonumber\\
S_\perp&=&-1.83\pm 1.04\stat\pm 0.23\syst.
\end{eqnarray}
The correlations between $C_+$ and $C_-$ and between $S_+$ and $S_-$
are $-0.46$ and $0.39$ respectively. All other correlations
are negligible.
Figure~\ref{fig:datafit_cp} shows the $\Delta t$ distributions and
asymmetry in yield between $B^0$ and $\Bzb$ tags, overlaid with the
result of the likelihood fit.  Because $R_\perp$ is small,
we have rather large statistical uncertainties for the measured
$C_\perp$ and $S_\perp$ values. We repeat the fit assuming that both
\CP-even and \CP-odd states have the same \CP asymmetry, i.e.\
$C_+=C_\perp=C$ and $S_+=S_\perp=S$.  We find
\begin{eqnarray}
C&=&-0.02\pm 0.11\stat\pm 0.02\syst,\nonumber\\
S&=&-0.66\pm 0.19\stat\pm 0.04\syst.
\end{eqnarray}
In both cases, the effective \CP asymmetries in the background are
found to be consistent with zero. To further test the consistency of the 
fitting procedure, the same analysis is applied to the 
$\Bz\to\Dss\Dstarm$ control sample. The result is consistent with
no \CP violation as expected.

The sources of systematic uncertainty on the \CP asymmetries and 
their estimated magnitudes are summarized in Table~\ref{tab:systematics}.  
We vary the yield and CP asymmetries of possible
backgrounds that peak under the signal.
We also vary fixed parameters in the fit for the
assumed parameterization of the $\Delta t$ resolution function, the
possible differences between the $B_{\rm flav}$ and $B_{\CP}$ mistag
fractions, and knowledge of the event-by-event beam-spot position.  We
evaluate the uncertainty due to possible interference between the
suppressed $b\to u\bar{c} d$ amplitude and the favored $b\to
c\bar{u}d$ amplitude for some tag-side decays~\cite{Long:2003wq}.  We
also include systematic uncertainties incurred from the finite MC
sample used to verify the fitting method. All of the systematic
uncertainties are much smaller than the statistical uncertainties.
\begin{table*}[htb]
\begin{ruledtabular}
\begin{tabular}{lcccccc}
Source  &  $C_+$ & $S_+$ & $C_\perp$ &
$S_\perp$ & $C$ & $S$
\\ \hline
Peaking backgrounds                  & 0.008 & 0.028 & 0.037 & 0.110 & 0.003 & 0.028  \\
\deltat\ resolution parameterization & 0.009 & 0.011 & 0.018 & 0.022 & 0.008 & 0.010 \\
Mistag fraction differences          & 0.008 & 0.024 & 0.016 & 0.035 & 0.008 & 0.024  \\
Beam-spot position                   & 0.004 & 0.007 & 0.019 & 0.042 & 0.003 & 0.005 \\
$\deltamd$, $\tau_B$                 & 0.004 & 0.006 & 0.016 & 0.004 & 0.001 & 0.006  \\
Angular resolution                   & 0.009 & 0.031 & 0.076 & 0.116 & 0.008 & 0.012  \\
Tag-side interference and others     & 0.014 & 0.009 & 0.017 & 0.021 & 0.014 & 0.009 \\
MC statistics                        & 0.005 & 0.013 & 0.031 & 0.150 & 0.001 & 0.013  \\
\hline
Total                                & 0.024 & 0.053 & 0.098 & 0.229 & 0.021 & 0.044 \\
\end{tabular}
\end{ruledtabular}
\caption{ Systematic errors on time-dependent \CP\ asymmetry
parameters for the decay \Bztodstdst.}
\label{tab:systematics}
\end{table*}

In summary, we have reported measurements of the \CP-odd fraction,
$R_\perp$, and time-dependent \CP asymmetries for the decay
\Bztodstdst.  The measurement is consistent with and supersedes the
previous \babar\ result~\cite{Aubert:2005rn}.  
The time-dependent
asymmetries are found to be consistent with the SM predictions.
The non-zero value of the measured $S_+$ indicates the evidence of 
CP violation at the $3.7\,\sigma$ confidence level.

We are grateful for the excellent luminosity and machine conditions
provided by our \pep2\ colleagues, and for the substantial dedicated
effort from the computing organizations that support \babar.  The
collaborating institutions wish to thank SLAC for its support and kind
hospitality.  This work is supported by DOE and NSF (USA), NSERC
(Canada), CEA and CNRS-IN2P3 (France), BMBF and DFG (Germany), INFN
(Italy), FOM (The Netherlands), NFR (Norway), MIST (Russia), MEC
(Spain), and STFC (United Kingdom).  Individuals have received support
from the Marie Curie EIF (European Union) and the A.~P.~Sloan
Foundation.


\begin{thebibliography}{99}
\bibitem{conjugate}
We imply charge conjugate modes throughout the paper.
\end{thebibliography}
\end{document}